%% file: template_peerReviewed.tex
\documentclass[convention,express-paper]{aesconf} 

\graphicspath{{./}{figures/}}

\usepackage[utf8]{inputenc}

\usepackage{microtype}
\usepackage{amsmath}
\usepackage{amsfonts}
\usepackage{amssymb}
\usepackage{amsthm}
\usepackage{graphicx}
\usepackage[numbers,square]{natbib}
\usepackage{setspace}
\edef\svtheparindent{\the\parindent}
\edef\svtheparskip{\the\parskip}

\usepackage{hyperref}

\newcommand\linesubsec[1]{{\sffamily\vspace{0.8mm}\noindent\textbf{#1 --- }}}

\usepackage[dvipsnames]{xcolor}
\newcommand\myshade{70}
\colorlet{mywholecolor}{MidnightBlue}
\hypersetup{
  linkcolor  = mywholecolor!\myshade!black,
  citecolor  = mywholecolor!\myshade!black,
  urlcolor   = mywholecolor!\myshade!black,
  colorlinks = true,
}

\makeatletter
 \DeclareRobustCommand\ref{%
    \@ifstar\@refstar\T@ref
  }%
  \DeclareRobustCommand\pageref{%
    \@ifstar\@pagerefstar\T@pageref
  }%
 \makeatother

\usepackage{booktabs}
\usepackage{url}
\usepackage{multirow}

\newcommand{\fref}[1]{Fig.~\ref{#1}}
\newcommand{\tref}[1]{Table~\ref{#1}}
\newcommand{\sref}[1]{Sec.~\ref{#1}}

\title{High-Fidelity Noise Reduction with Differentiable Signal Processing}

\author[1,2]{Christian J. Steinmetz}
\author[2]{Thomas Walther}
\author[1]{Joshua D. Reiss}

\affil[1]{Centre for Digital Music, Queen Mary University of London, UK}
\affil[2]{Tape It Music GmbH}

\correspondence{Christian J. Steinmetz}{c.j.steinmetz@qmul.ac.uk}

\lastnames{Steinmetz, Walther, Reiss}

\shorttitle{Differentiable noise reduction}

\expresspapernumber{}

\include{aestitle}

\begin{document}

\twocolumn[
\maketitle 

\begin{onecolabstract}
\noindent Noise reduction techniques based on deep learning have demonstrated impressive performance in enhancing the overall quality of recorded speech. While these approaches are highly performant, their application in audio engineering can be limited due to a number of factors. These include operation only on speech without support for music, lack of real-time capability, lack of interpretable control parameters, operation at lower sample rates, and a tendency to introduce artifacts. On the other hand, signal processing-based noise reduction algorithms offer fine-grained control and operation on a broad range of content, however, they often require manual operation to achieve the best results. To address the limitations of both approaches, in this work we introduce a method that leverages a signal processing-based denoiser that when combined with a neural network controller, enables fully automatic and high-fidelity noise reduction on both speech and music signals. We evaluate our proposed method with objective metrics and a perceptual listening test. Our evaluation reveals that speech enhancement models can be extended to music, however training the model to remove only stationary noise is critical. Furthermore, our proposed approach achieves performance on par with the deep learning models, while being significantly more efficient and introducing fewer artifacts in some cases. Listening examples are available online: {\footnotesize{\url{https://tape.it/research/denoiser}}} \vspace{0.22cm}

\end{onecolabstract}
]

\section{Introduction}\vspace{-0.1cm}

Traditional noise reduction techniques, such as the Wiener filter~\cite{wiener1964extrapolation}, spectral subtraction~\cite{boll1979suppression}, and spectral gating~\cite{hicks1996evolution} increase the signal-to-noise ratio by exploiting assumptions about the statistics of the source and noise signals.  
These approaches employ different methods for estimating the statistics of the noise and may make assumptions, such as noise stationarity, both of which can be a limiting factor of performance.
However, when operated by experienced users, commercial tools based on these methods, such as iZotope RX Spectral Denoise, are capable of transparent noise reduction for diverse signal types in many scenarios.

Recently, deep learning approaches have demonstrated superior performance in the joint task of noise reduction and de-reverberation of speech, often referred to as speech enhancement~\cite{yuliani2021speech}. 
These approaches overcome the limitations of signal processing techniques since they use data to train large estimators that make fewer assumptions. 
However, speech enhancement models cannot be used directly to enhance the quality of other sources, such as musical instruments, since they have been trained to extract or generate only speech. 
As a result, enhancing music signals with speech enhancement models often results in the removal of musical sources or the introduction of corruptions, such as transforming instruments into pseudo-speech utterances or changing the pitch of vocals\footnote{\vspace{-0.1cm}Examples at \url{https://tape.it/research/denoiser}}. 

This motivates the related task of music enhancement, which has so far been less studied. 
Recent works have considered adapting the training objective for speech enhancement models to focus on music signals~\cite{kandpal2022music, moliner2022two, chae2023exploiting}. 
While these approaches can enhance music signals, they suffer from many of the same limitations of existing deep learning-based speech enhancement systems, namely operation at lower sample rates, the introduction of artifacts, high compute cost, and lack of parametric control. 
These limit the applicability of such algorithms in the context of audio engineering.

In this work we provide a number of contributions. 
Firstly, like previous work, we extend speech enhancement models for noise reduction of music signals. 
However, we find standard enhancement pipelines that include removal of both stationary and non-stationary noise problematic. 
Our results indicate these pipelines produce models with more artifacts as compared to models trained to remove only stationary noise.
Secondly, we propose a hybrid signal processing and deep learning approach that is capable of full-band noise reduction for stereo signals consisting of speech, music, and general audio. 
Unlike previous hybrid approaches, our method utilizes differentiable signal processing to train with the denoiser in the loop.
Thirdly, in the design of our hybrid approach, we demonstrate that existing differentiable approximations of dynamic range processor ballistics are problematic. 
We circumvent this using gradient approximation schemes, which we make scalable with a two-stage training process. 

We evaluate our approach compared to strong baselines using both objective metrics and a listening test. 
While our adapted speech enhancement models bring the best performance in objective metrics, the listening test reveals that our hybrid approach achieves performance on par with large deep learning models as well as commercial, signal processing-based noise reduction systems. 
We achieve this while enabling fully automatic operation, yet also providing parametric control and requiring an order of magnitude less compute.

\section{Related Work}\vspace{-0.1cm}
\linesubsec{Speech enhancement}
While many deep learning speech enhancement systems consider only narrow band content, recent approaches such as HiFi-GAN~\cite{su2020hifi} and DeepFilterNet~\cite{schroter2022deepfilternet} have demonstrated success in enhancing full-band speech at sample rates up to 44.1\,kHz, enabling applications in audio engineering\footnote{\url{https://podcast.adobe.com/enhance}}. 
However, these approaches do not work for non-speech recordings, require considerable compute, introduce artifacts in challenging scenarios, and do not offer controllable enhancement.
To address these limitations we combine traditional signal processing noise reduction techniques with a neural network controller in order to achieve high-fidelity noise reduction on diverse content.

\linesubsec{Hybrid approaches} 
Our work is not the first to consider combining signal processing noise reduction techniques with deep learning. 
Highly related are approaches like RNNoise~\cite{valin2018hybrid}, and its successor PercepNet~\cite{valin2020perceptually}, which use perceptually motivated signal processing components. 
However, they are designed to exploit characteristics of speech signals, limiting their applicability to non-speech signals. 
In addition, these approaches are subject to introducing noticeable time-varying artifacts and are without controls that enable users to limit distortions caused by the noise reduction process.
Furthermore, these approaches are trained to regress ground truth parameter values which may not produce perceptually optimal results. 
Our approach overcomes these challenges by leveraging a more general noise reduction framework that instead focuses on removing largely stationary noise, which is trained end-to-end with the denoiser in the loop. 

\linesubsec{Music enhancement} 
Similar to speech enhancement, music enhancement generally entails reduction of stationary and non-stationary noise as well as de-reverberation. 
Recent works have considered extending speech enhancement models for this task. 
Approaches include a two stage method that enhances spectral representations and then uses a diffusion vocoder~\cite{kandpal2022music}, adaptation of the Conformer architecture from speech separation~\cite{chae2023exploiting}, and diffusion models operating on complex spectral representations~\cite{moliner2022two, irigaray2023noise}.
In this work, we consider only noise reduction, with a focus on the reduction of largely stationary noise.
We do not consider reducing the effect of room reflections or reverberation, making comparison to these works less relevant.

\section{Method}

The noisy signal model in the time domain is given by 
\begin{equation}
    x(n) = y(n) + w(n)
\end{equation}
where $x(n)$ is the noisy signal, $y(n)$ is the original signal, and $w(n)$ is the additive noise with $n$ as the discrete time index. 
In the noise reduction task, our goal is to produce an estimate of the clean signal $\hat{y}(n)$ that is perceptually similar to the original clean signal. 

\begin{figure}[t]
    \centering
    \includegraphics[width=\linewidth]{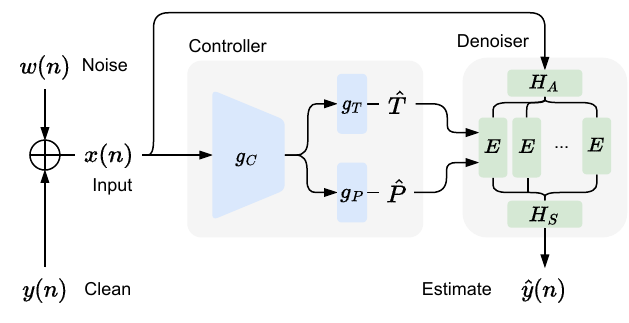}
    \caption{High-level view of our noise reduction system.}
    \label{fig:overall}
    \vspace{-0.2cm}
\end{figure}

As shown in \fref{fig:overall}, our noise reduction system is composed of two subsystems: a signal processing denoiser and a neural network controller, which is trained to estimate time-varying controls for the denoiser.
It is composed of a main network $g_C$ featuring a convolutional recurrent architecture, as well as two prediction networks $g_T$ and $g_P$, which estimate the spectrum of the noise $\hat{T}(b)$ across $B$ bands, as well as the denoiser control parameters $\hat{P}$.
The denoiser employs spectral gating~\cite{hicks1996evolution}, using a set of filterbanks and respective dynamic range expanders in order to reduce the audibility of noise across different frequency bands of the input.

Unlike recent speech enhancement models that use a large neural network to both analyze and process the noisy audio signal, our approach uses a neural network to first analyze the signal and estimate control parameters, while the audio signal is processed using only the signal processing denoiser. 
This enables improved efficiency at high sample rates as well as interpretable control parameters that can be used to fine-tune the noise reduction. 
Furthermore, this enables us to restrict the minimum and maximum values of certain parameters, such as the attack and release times, which ensures that even in challenging scenarios our denoiser will not act too aggressive, helping to avoid artifacts.

\begin{figure}[t!]
    \centering
    \includegraphics[width=\linewidth]{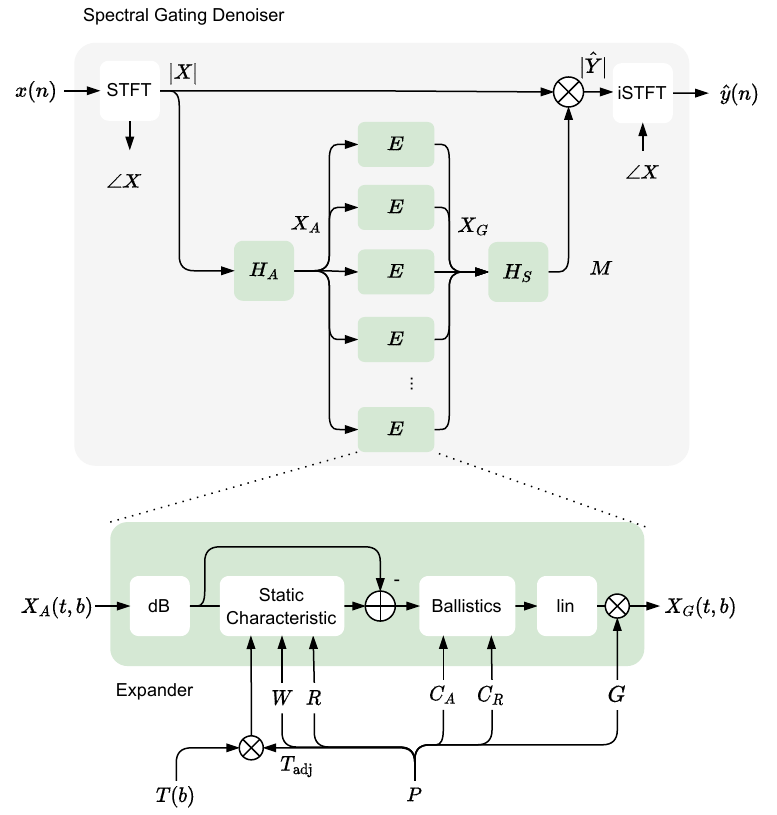}
    \caption{Multi-band spectral expander architecture.}
    \label{fig:expander}
    \vspace{-0.2cm}
\end{figure}

\subsection{Signal processing denoiser}\label{sec:denoiser}

We use a denoiser based upon spectral gating, also known as spectral expansion~\cite{moorer1986linear, hicks1996evolution}.
Unlike spectral subtraction, where each time frame is treated the same by subtracting a constant energy from the each magnitude frequency bin~\cite{boll1979suppression}, spectral expansion applies a time-varying gain reduction across a set of frequency bands as a function of the energy in each band. 
We opt for this approach since spectral subtraction is known to introduce musical noise artifacts~\cite{lukin2007suppression}, and our initial testing indicated that spectral gating produced less artifacts when optimal control parameters were selected.

Our multi-band spectral expander is composed of a set of filterbanks and dedicated band-wise expanders, as shown in \fref{fig:expander}.  
To carry out processing in the frequency domain the short-time Fourier transform (STFT) is employed, splitting the time domain signal into overlapping windows with hop size $H$, and taking the DFT of each window with FFT size $N$. 
This produces a frequency domain representation of the input $X(t, f)$
where $t$ corresponds to the time frame index and $f$ corresponds to the frequency index. 
The denoiser operates on the magnitude spectrogram $|X| \in \mathbb{R}^{S \times F}$ where $F$ is the number of frequency bins and $S$ is the number of time frames in the sequence. 
After transforming the input signal to the frequency domain, the next step is to separate the magnitude spectrogram into a set of $B$ perceptually spaced frequency bands. This is achieved with the analysis filterbank, which is given by a matrix $H_A \in \mathbb{R}^{F \times B}$, and is used to produce $ X_A = |X|^T H_A$, a representation of the energy within each band.

Next, an independent dynamic range expander operates on each frequency band. 
The design follows from the digital dynamic range compressor~\cite{giannoulis2012digital}.
The signal in each band is first converted to the log domain 
and then fed to the gain computer.
The gain computer applies the static gain characteristic across each frequency band $b$, attenuating the signal below the individual threshold $T(b)$ defined for each band with the ratio $R$ and the knee width $W$.
We opt for a soft knee, which is defined by
\begin{equation} \footnotesize
    X_L(t,b) = 
    \begin{cases}
        X_a(t,b) & X_a(t,b)< \left( T\left(b\right) - \frac{W}{2} \right) \\
        X_{L_{\text{at}}}(t,b) & \left( T(b) - \frac{W}{2} \right) \leq X_a(t,b) \leq \left( T(b) + \frac{W}{2} \right) \\
        X_{L_{\text{below}}}(t,b) & X_a(t,b) > \left( T(b) + \frac{W}{2} \right) 
    \end{cases},
\end{equation}
where the reduction at the knee is given by
\begin{equation}
    X_{L_{\text{at}}(t,b)} = X_a(t,b) + \frac{\left(1 - R\right) \left( X_a(t,b) - T(b) - \frac{W}{2} \right)^2}{2W},
\end{equation}
and the reduction below the knee is given by
\begin{equation}
    X_{L_{\text{below}}}(t,b) = T(b) + \left( X_a\left(t,b\right) - T\left(b\right) \right) \times R.
\end{equation}

While it is possible to use a unique ratio and knee width for each band, we opt to use the same ratio $R$ and knee width $W$ for each band. Only the threshold is unique to each band. 
The gain reduction in each band and frame $t$ is given by $X_G(t,b) = X_{L}(t,b) - X_A(t,b).$

Afterwards a smoothing peak detector is applied, which produces the characteristic ballistics in dynamic range processors. 
This detector is implemented as a branching first-order recursive filter where the two filters have time constants $\alpha_A$ and $\alpha_R$ given by the attack and release times, $C_A$ and $C_R$, where $f_s$ is the audio sample rate and $H$ is the hop size
\begin{equation}
\begin{split}
\alpha_A  = \exp \left( \frac{-\log \left(9\right)}{\frac{f_s}{H} \times C_A} \right) , \
\alpha_R  = \exp \left( \frac{-\log \left(9\right)}{\frac{f_s}{H} \times C_R} \right).
\end{split}
\end{equation}
The output of the smooth peak detector is
\begin{equation}
{\footnotesize
    X_{G_s}(t) = 
    \begin{cases}
        \alpha_A X_{G_s}(t-1) + \left(1 - \alpha_A\right)X_G(t) & X_G(t) \leq X_{G_s}(t-1) \\
        \alpha_R X_{G_s}(t-1) + \left(1 - \alpha_R\right)X_G(t) & X_G(t) > X_{G_s}(t-1)
    \end{cases}.
    }
\end{equation}
We drop the $b$ index for simplification. 
Finally, we convert the gain reduction values from the log-domain back to the linear domain.

Once the linear gain reduction is computed for each band, given by $X_{G_s}$, we use the synthesis filterbank $H_s$ to project the $B$ bands back to the $F$ frequency bins of the noisy magnitude spectrogram. 
Then, this mask $M \in \mathbb{R}^{S \times F}$ is applied via point-wise multiplication with the noisy magnitude spectrogram $\hat{|Y|} = |X| \odot (G \cdot X_{G_s})$
where $G$ is the gain adjustment in linear domain. 
Then the signal is converted back to the time domain using the inverse STFT and the original noisy phase.

While we introduce here a monophonic formulation, we extend the denoiser to the multi-channel case in one of two ways. 
The simplest extension is a dual mono formulation where each input channel features a separate set of expanders. 
However, this can lead to perceivable distortions in the stereo image if significant gain reduction is applied to one channel. 
To address this we can also operate the multi-channel denoiser in a ``linked'' stereo mode where the energy in each band is the average between both channels, which can then be used as the side-chain signal to determine a shared gain reduction applied equally to both channels.

While initially our approach may appear distinct from existing deep learning approaches, spectral gating is conceptually related to deep learning approaches for speech enhancement that estimate a time-varying multiplicative mask for each frame of the noisy magnitude spectrogram~\cite{yuliani2021speech}. 
However, unlike the deep learning approaches, the process of estimating the mask in our approach is based on a set of interpretable equations that dictate how fast and by how much the signal should be attenuated in each frequency band.
This is key to our approach as we are able to restrict the ranges of parameters to ensure that even in worst case scenarios the operation of the denoiser is not overly aggressive.

\begin{figure*}[t]
    \vspace{-0.25cm}
    \centering
    \includegraphics[width=0.95\linewidth]{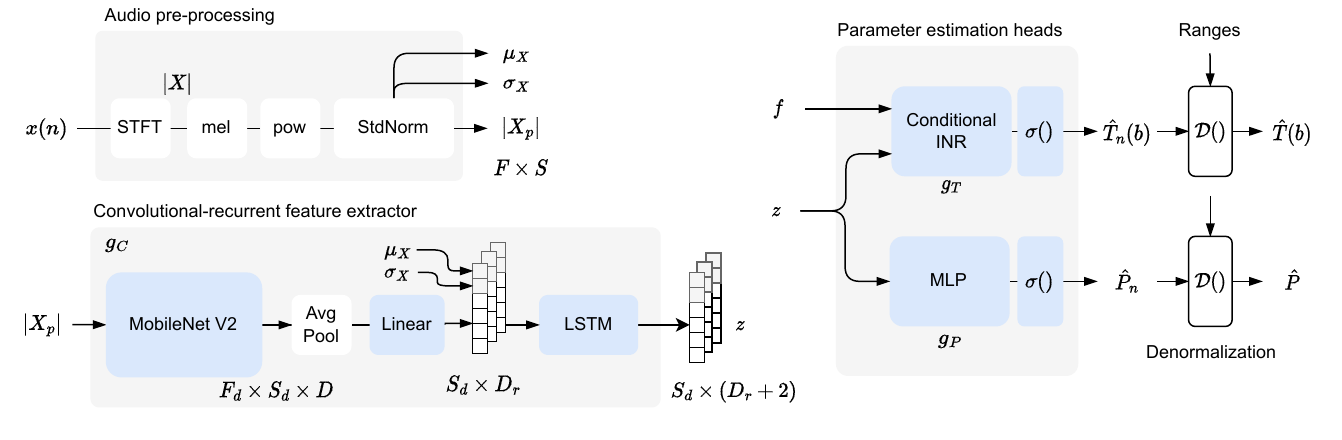} 
    \vspace{-0.1cm}
    \caption{Controller network with pre-processing, feature extractor, and parameter estimation heads.}
    \label{fig:controller}
    \vspace{-0.25cm}
\end{figure*}

\subsection{Neural network controller}

The neural network controller is composed of three major components: audio pre-processing, a convolutional-recurrent feature extractor, and specialized parameter estimation heads, as shown in \fref{fig:controller}.
The controller aims to extract relevant features from the noisy input signal $x(n)$ to estimate both the spectrum of the noise across the denoiser frequency bands $\hat{T}(b)$ as well as the optimal denoiser parameters $\hat{P} = \{C_A, C_R, W, R, G\}$ to effectively reduce the stationary noise component while introducing minimal artifacts. 

The audio pre-processing pipeline begins by first transforming the time domain signal via the STFT. 
This is followed by projection onto the mel-basis and exponential scaling with $|X|^{\beta}$. 
Before passing the spectrogram to the feature extractor we apply StdNorm, 
$|X_p| = \frac{|X| - \mu_X}{\sigma_X}$,
which normalizes the magnitude spectrogram by its mean $\mu_X$ and standard deviation $\sigma_X$.
    
The main component of the controller is the MobileNetV3~\cite{howard2019searching}, which we have adapted by removing all BatchNorm layers, and modified so the input layer operates on magnitude spectrograms as images with one channel.
The output of the MobileNetV3 is a 3-dim downsampled feature map of shape $F_d \times S_d \times D$, 
where $F_d$ is the downsampled frequency dimension, $S_d$ is the downsampled sequence length, and $D$ is the feature dimension.
We apply average pooling across the frequency dimension to produce a 2-dim representation of size $S_d \times D$.
This representation is then passed through a linear layer, shared across the temporal dimension, to reduce the feature dimension to $D_r$.
To provide information about the absolute scale of the input we concatenate the frame-wise $\mu_X$ and $\sigma_X$ to this representation and then pass it through a LSTM to produce the final sequence of latent representations $z \in \mathbb{R}^{D_r \times S}$.

The final components of the controller are two specialized parameter estimation heads. 
The first network $g_P$ is tasked with predicting the denoiser parameters $\hat{P}$ and is constructed as a simple 3-layer MLP, which operates on the latent representation $z$. 
The second network $g_T$ is tasked with estimating the energy of the noise in each frequency band of the denoiser $T(b)$.
This network is implemented as a conditional implicit neural representation (INR)~\cite{sitzmann2020implicit}, which features a combination of linear layers with sinusoidal activation functions, along with a modulator network. 
The modulator network produces scaling values for the intermediate representations of the INR based upon $z$, while $f$ represents a set of frequency band indices on $[-1,1]$ for which the network is estimating the energy of the noise spectrum. 
The network architecture is motivated by the fact that the noise spectrum can itself be modeled as a continuous signal across the frequency range, and was found to perform better than a simple MLP.

To stabilize training we apply a sigmoid activation to the output of $g_T$ and $g_P$. 
This scales all parameters between 0 and 1. 
Then, in order to rescale the parameters into ranges appropriate for the denoiser we apply a denormalization step $\mathcal{D}()$.
This operation individually rescales each parameter to a predefined range. 
We select ranges through initial testing to ensure the denoiser does not have the ability to operate too aggressively, yet is still capable of reducing noise.
In addition, the controller will generate control parameters for each input segment of length $L$ samples. 
We then produce an upsampled sequence of control parameters that changes at every STFT frame by linearly interpolating between the control parameter values for each segment.

\subsection{Differentiable training}

Since we do not have a priori the optimal denoiser parameters, to facilitate training of the controller network we must backpropagate through the denoiser by computing the error between the output of the denoiser $\hat{y}(n)$ and the ground truth clean signal $y(n)$.
This can be enabled by differentiable signal processing~\cite{hayes2023review}.
Predominant approaches for differentiable signal processing include explicit automatic differentiation~\cite{engel2020ddsp}, neural proxies~\cite{steinmetz2021automatic, steinmetz2022style}, and gradient approximation~\cite{ramirez2021differentiable}. 

\begin{figure}[t]
    \centering
        \vspace{-0.2cm}
    \includegraphics[width=0.9\linewidth]{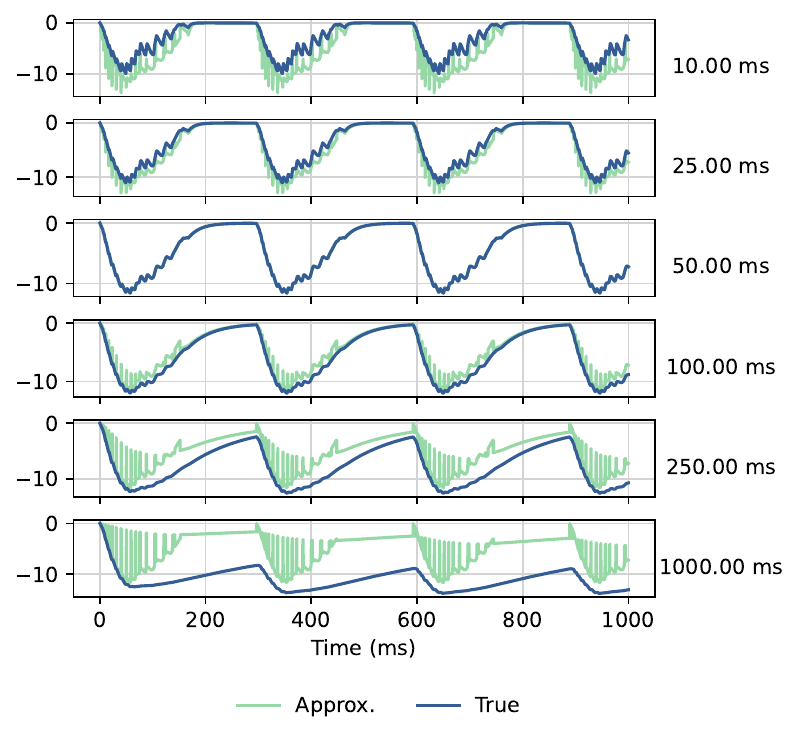}
    \vspace{-0.5cm}
    \caption{Gain reduction produced by a compressor with recursive and approximate ballistics.}
    \label{fig:ballisitics}
    \vspace{-0.5cm}
\end{figure}

The design of an automatic differentiation dynamic range processor is problematic due to the ballistics, which implement switching or branching behavior as described in \sref{sec:denoiser}. 
While this branching is not exactly differentiable it can be implemented in an approximately differentiable manner.
However, this formulation is generally not practical due to backpropagation through time at audio sample rate~\cite{wright2022grey}.
This motivates simplified compressor designs with a single time constant~\cite{steinmetz2022style}, or approximations of the behavior without the recursion using two independent filters~\cite{colonel2022approximating}. 

However, our investigations revealed that these approaches are not sufficient to capture the behavior of the ballistics when the attack and release times differ significantly.
In \fref{fig:ballisitics} we plot the gain reduction at the output of a simple dynamic range compressor when using the true switching ballistics and the previously proposed approximation~\cite{colonel2022approximating}. 
An attack time of 50\,ms is used and the release time is varied from 10\,ms to 1000 \,ms. 
When the difference between the attack and release are small the approximation is close to the true ballistics. 
However, as the release time increases the difference between the curves becomes significant, especially at a release time of 250\,ms and beyond.

\begin{figure}[t!]
    \centering
    \vspace{-0.2cm}
    \includegraphics[width=0.9\linewidth]{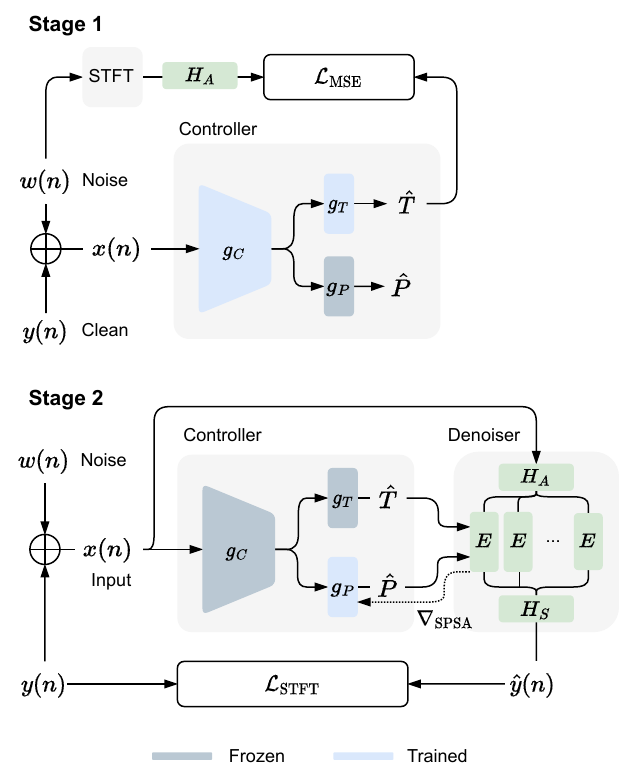}
    \caption{Two stage training process. }
    \vspace{-0.47cm}
    \label{fig:training}
\end{figure}

While it would be possible to completely avoid backpropagation through the denoiser by simply training the controller computing a loss on the output of $g_T$ using the ground truth noise spectrum $T(b)$, we found this results in suboptimal performance. 
Instead, adjustment of the noise thresholds and the other denoiser parameters is required.
Therefore, to facilitate training we opt to use gradient approximation~\cite{ramirez2021differentiable}. 
Leveraging stochastic simultaneous perturbation approximation (SPSA)~\cite{spall1992multivariate} provides a more scalable approach than finite differences (FD), however, we found training our system in this manner both slow and prone to instability, hindering performance.
To address this, we introduced a two-stage training process as shown in \fref{fig:training}.

The first stage involves pre-training the feature extractor $g_C$ and the noise spectrum estimation network $g_T$ in a supervised task for estimation of the noise spectrum $T(b)$. 
To achieve this, we compute the ground truth noise spectrum by transforming the noise signal $w(n)$ to the frequency domain with the STFT and then pass this signal through the analysis filterbank $H_A$ taking the mean across time frames. 
This enables us to compute the mean squared error between the estimated spectrum and ground truth during training.

Then, during the second stage, the weights of $g_C$ and $g_T$ are frozen and only the parameter estimation network $g_P$, a small MLP, is trained using gradients from the approximation method.
Here we compute the loss between the denoised audio at the output of the denoiser $\hat{y}(n)$ and the ground truth clean signal $y(n)$, using the multi-resolution STFT loss~\cite{yamamoto2020parallel}.
This two-stage approach stabilizes training and reduces overall training time since we can use most of the gradient steps to update the weights of $g_C$, which contains the majority of the weights. 
However, this method enables the controller to adapt its estimation of the denoiser parameters based on features from the noisy input.

\setlength{\tabcolsep}{3.0pt}
\renewcommand{\arraystretch}{0.85}
\begin{table*}[]
    \vspace{-0.3cm}
    \centering
    \begin{tabular}{l c c c c c c c c c c c c c c}\toprule
        Approach        &  \multicolumn{3}{c}{VCTK} & \multicolumn{3}{c}{DSD100} & \multicolumn{3}{c}{GuitarSet} & \multicolumn{3}{c}{VocalSet} \\ \cmidrule(lr){2-4} \cmidrule(lr){5-7} \cmidrule(lr){8-10} \cmidrule(lr){11-13}
                            & SI-SDR & STFT & FAD & SI-SDR  & STFT & FAD & SI-SDR  & STFT & FAD & SI-SDR & STFT & FAD   \\ \midrule
        Input               & 27.9  & 0.518 & 0.112 & 28.99 & 0.488 & 0.067 & 28.33 & 0.491 & 0.087 & 27.4 & 0.829 & 0.148  \\ \midrule
        noisereduce         & 5.73  & 1.571 & 0.118 & 6.39  & 1.678 & 0.151 & 6.62 & 1.499 & 0.092 & 8.09 & 1.407 & 0.117 \\ 
        RNNoise             & 12.7  & 0.606 & 0.089 & 1.80  & 1.400 & 0.187 & 8.77 & 0.792 & 0.099 & 6.58 & 1.115 & 0.181 \\ \midrule
        HDemucs             & 29.9 & 0.208 & 0.032 & 29.63 & 0.192 & 0.016 & 30.04 & 0.207 & 0.028  & 34.5 & 0.307 & 0.038 \\
        HDemucs (DNS)       & 29.2 & 0.236 & 0.036 & 28.84 & 0.246 & 0.022 & 29.01 & 0.240 & 0.037 & 32.6 & 0.349 & 0.044  \\
        DCUNet              & 27.26 & 0.242 & 0.032 & 24.46 & 0.246 & 0.025 & 25.42 & 0.237 & 0.028 & 28.9 & 0.375 & 0.034 \\ \midrule
        Tape It             & 28.72 & 0.406 & 0.067 & 29.03 & 0.388 & 0.034 & 29.01 & 0.370 & 0.050 & 30.0 & 0.614 & 0.079 \\
        Tape It (Stage 1)   & 26.98 & 0.459 & 0.064 & 28.42 & 0.418 & 0.035 & 28.34 & 0.409 & 0.046 & 30.1 & 0.621 & 0.068 \\
        \bottomrule
    \end{tabular}
    \caption{Models evaluated across SI-SDR ($\uparrow$), mel STFT error ($\downarrow$), and FAD ($\downarrow$) on held out test data.}
    \vspace{-0.4cm}
    \label{tab:metrics-by-source}
\end{table*}


\newpage
\section{Experiments}

To train our noise reduction models we construct noisy examples by combining noise-free sources with largely stationary noises. 
To facilitate generalization and expose the models to a wide range of sources we combine a number of existing datasets in our synthetic data generation pipeline. 
We include VCTK~\cite{yamagishi2019cstr} for speech, GuitarSet~\cite{xi2018guitarset} for acoustic guitar, VocalSet~\cite{wilkins2018vocalset} for vocals, and DSD100~\cite{dsd100} for general instrumentation. 
We source noise from three datasets. These include purely synthetic noises generated by filtering white noise, recordings collected from Freesound, along with a set of recordings we collected using mobile devices. 

During training we randomly sample audio segments of 262144 samples at $f_s = 44.1$\,kHz ($\approx$ 6 sec), one from a source dataset, and one from a noise dataset.
To increase robustness to room reverberation, we convolve at random either the source, or the linear combination of the noise and source, with a randomly sampled impulse response. 
We source impulse responses from the MIT IR Survey~\cite{traer2016statistics} and the EchoThief Impulse Response Library\footnote{\vspace{-0.2cm}\url{http://www.echothief.com}}.
We also apply augmentations including time stretching, pitch shifting, MP3 compression, time-varying gain, and random filtering.
The source and noise signals are combined with random scaling factors such that the relative difference in perceptual loudness~\cite{steinmetz2021pyloudnorm} is sampled uniformly from -48 to -12 dB. 

To validate our approach we compare against a set of strong baselines with both objective metrics and a perceptual listening test. We consider a comparable hybrid approach, RNNoise~\cite{valin2018hybrid}, and a traditional signal processing method, noisereduce~\cite{sainburg2020finding}, as baselines.
To form stronger baselines, we also adapt existing speech enhancement models, Hybrid Demucs (HDemucs)~\cite{defossez2021hybrid} and DCUNet~\cite{choi2018phase}, which we train using our data generation pipeline.
We consider iZotope RX Spectral Denoise as a commercial system, however it is included only in the perceptual evaluation since there is not a scalable way to run it across the test set.

We conduct two further experiments. 
First, to investigate the effect of directly adapting existing speech enhancement pipelines to our general noise reduction task, we train a variant of HDemucs using the same four source datasets, but instead source noise from DNS, an existing large-scale noise dataset~\cite{dubey2022icassp}. 
This enables us to compare the impact of training the model to remove both stationary and non-stationary noise.
Second, we investigate the benefit of our two-stage training approach by using a variant of our model that uses only the first stage of training and fixed denoiser parameters.

Our denoiser uses $B=27$ perceptually spaced bands based on the bark scale. Both the denoiser and the controller use FFT size $N=1024$ with hop size $H=256$.
All models are trained with AdamW and an initial learning rate of $1\cdot 10 ^{-4}$ for a total of 200\,k steps, reducing the learning rate by a factor of 10 at 80\% and 95\% through training. 
For our two stage training process, we first train noise spectrum estimation for 150\,k steps and train the remaining 50\,k steps using gradient approximation with $\epsilon=0.01$.
 We limit the noise thresholds from -80 to 24\,dB, attack from 10 to 1000\,ms, release from 50 to 250\,ms, threshold adjustment from -12 to 32\,dB, makeup gain from -12 to 12\,dB, ratio from 2 to 10, and knee from 0 to 24\,dB. 
For the multi-resolution STFT loss we use \texttt{auraloss}~\cite{steinmetz2020auraloss} with window sizes of $256, 1024, 4096, 16384$ and hop sizes of 1/2 using a linear combination of the linear and log spectrograms.
We use clip the magnitude of gradients to 4.
We set the batch size to the largest value that will fit with a single 16GB T4 GPU, enabling DCUNet with a batch size of 16, HDemucs with 4, and our model with 32.

\begin{figure*}[!t]
    \vspace{-0.5cm}
    \centering
    \includegraphics[width=\linewidth]{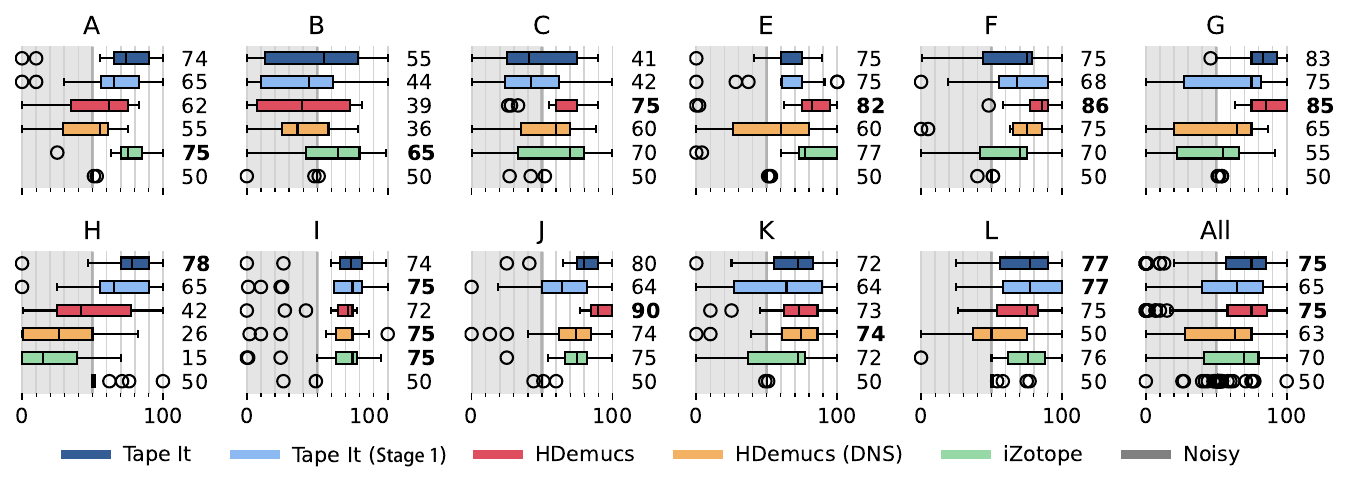}
    \vspace{-0.6cm}
    \caption{Scores across the eleven stimuli. The median values are denoted for each method on the right with the best performing method shown in boldface. Scores below 50 denote lower preference as compared to the noisy input, which is shaded as the grey region in each plot. ``All'' denotes aggregated scores across all stimuli.}
    \vspace{-0.25cm}
    \label{fig:preference-boxplot}
\end{figure*}

\section{Results}

\subsection{Objective evaluation}

We report SI-SDR, FAD~\cite{kilgour2019frechet}, and the mel-STFT error in \tref{tab:metrics-by-source} across 1000 examples from held out test data using four source datasets.
We compute FAD using CLAP~\cite{wu2023large}, which has been trained on music at $f_s =48$\,kHz, and the mel-scaled STFT using \texttt{auraloss}~\cite{steinmetz2020auraloss} with  FFT size 2048 and hop size 512.
The traditional signal processing baseline, noisereduce, performs poorly across all metrics and datasets. 
RNNoise performs somewhat better, however it does not bring an improvement over the input when considering the full reference metrics, only in FAD when considering speech. As expected, FAD is significantly higher for the non-speech sources, since RNNoise was trained only on speech. 
All models we trained bring an improvement compared to the input on all metrics and across both speech and music.
HDemucs achieves superior performance in the full reference metrics across all datasets, however, DCUNet outperforms in FAD. 
Tape It performs better than other baselines but worse compared to HDemucs and DCUNet in all metrics. 

From our ablation study we find that training HDemucs with noise from DNS dataset does result in a small decrease in SI-SDR, and a somewhat more significant decrease in STFT as well as FAD performance, indicating potential downsides of this training pipeline. 
When comparing our approach using only the Stage 1 training we observe noticeable decrease in performance for SI-SDR and STFT, yet FAD is improved by a small margin on VTCK, GuitarSet, and VocalSet. 
We hypothesize this result is due to more aggressive noise reduction in the model with only Stage 1 training, however, from our listening this comes at the cost of distortion to the source and a loss of high frequencies.

\renewcommand{\arraystretch}{0.9}
\begin{table}[]
\centering
    \begin{tabular}{c l}
        \toprule
        ID & Source \\ \midrule
        A &	    Female speech \\
        B & 	Organ in cathedral \\
        C & 	Female vocal \\
        D & 	Electric guitar* (omitted) \\
        E & 	Bird calls \\
        F & 	Acoustic guitar (high noise) \\
        G & 	Acoustic guitar and male vocal \\
        H & 	Jazz ensemble with trumpet \\
        I & 	Solo piano \\
        J & 	Male speech \\
        K & 	Ukulele \\
        L & 	Male vocal \\
        \bottomrule
    \end{tabular}
    \caption{Listening test stimuli.}
    \vspace{-0.5cm}
    \label{tab:stimuli}
\end{table}

\subsection{Perceptual evaluation}\label{sec:perceptual}

To better understand the noise reduction performance we designed a perceptual evaluation using the Go Listen platform~\cite{barry2021go}. 
To evaluate the systems we selected twelve realistic recordings containing audible, yet largely stationary background noises, covering a range of content types, as shown in \tref{tab:stimuli}.
A total of 26 participants who self-reported experience in critical listening and have no known hearing impairments were enlisted to complete the evaluation.
We include four methods in the evaluation: Tape It, Tape It (only Stage 1 training), HDemucs, HDemucs (DNS), and iZotope RX Spectral Denoise. 
Participants were presented with a noisy recording, which was marked as the reference, along with five other stimuli: the four methods, and a hidden version of the noisy recording, which they were instructed to give a score of 50.
Participants were asked to provide a score from 50 to 100 reflecting how much they preferred a recording compared to the reference, or to provide a score from 0 to 50 reflecting how much a method harmed performance, with 0 being the worst. 

We performed post-filtering, removing responses from participants that rated the hidden reference more than $\pm 10$ away from the target score 50 in more than 2 recordings, leaving responses from 18 participants.
We found many participants were unable to easily detect the noise in stimulus D, so we omitted it from our analysis.
Scores across the eleven stimuli shown in \fref{fig:preference-boxplot} with stimuli described in \tref{tab:stimuli}.
We denote the region below 50 in gray to make clear that scores in this area indicate participants felt the method \emph{harmed} performance compared to the unprocessed noisy recording. 

We find significant variance in performance among the methods and between recordings. 
For example, HDemucs significantly outperforms other approaches in J and E, while Tape It shows strong performance in H and L where other methods struggle. 
When looking at the aggregate scores (All) we see that the median scores for Tape It and HDemucs are equal at 75, followed by iZotope at 70. Tape It with Stage 1 training (65) and HDemucs (DNS) (63) perform noticeably worse. 

To formalize these findings we first perform the Kruskal-Wallis test and find a significant difference among the medians ($p = 1.60 \cdot 10^{-33}$). We then apply Dunn’s test performing all pairwise comparisons of the medians using the Bonferroni correction. 
First, we find that the differences in median scores of Tape It and HDemucs (DNS) and the noisy input signal are significant $(p_{\text{adj}} = 0.0004)$ and $(p_{\text{adj}} = 4.06 \cdot 10^{-22})$.
However, we do not find the difference in the medians significant for Tape It compared to HDemucs or iZotope. 
Our informal listening points to differences in these methods, however, preference for these approaches may be context, source, and user dependent.

\subsection{Efficiency}

We report the real-time factor of our proposed method and baselines in \tref{tab:runtime}. 
The real-time factor is how much faster than real-time audio is processed by each method using an input of stereo audio at $f_s=44.1$\,kHz, 12 seconds in duration. 
Timings are averaged across 100 runs on both CPU and GPU on a machine using a Intel(R) Xeon(R) Platinum 8259CL CPU @ 2.50GHz and a NVIDIA Tesla T4 16GB GPU.
In order to run baselines that do not support streaming out of the box, such as DCUNet and HDemucs, we compute the output from the models in frames of 262144 samples and apply overlap-add processing with 50\% overlap using a Hann window.
Our proposed method is able to achieve a real-time factor of 26.8, nearly 15 times faster than HDemucs when running inference on CPU.
RNNoise achieves the best performance, which is due to a combination of its small parameter count, as well as the optimized C++ implementation. Note that the other models run in PyTorch. 
As a result, implementation optimizations in our approach could therefore lead to comparable or superior performance.

\setlength{\tabcolsep}{5.0pt}
\begin{table}[]
    \centering
    \begin{tabular}{l c c c} \toprule
         Approach       & Params & \multicolumn{2}{c}{Real-time factor}    \\ \cmidrule(lr){3-4}
                        &          & CPU  & GPU           \\ \midrule
         DCUNet         &  2.37\,M & 2.14 & 12.8        \\
         HDemucs        &  83.6\,M & 1.80 & 10.7 \\ 
         noisereduce    & -        & 25.2 & - \\ 
         RNNoise        &  87.5\,k & \textbf{39.9} & - \\ \midrule
         Tape It        &  4.57\,M &  26.8 & \textbf{44.9}  \\
         \bottomrule
    \end{tabular}
    \caption{Real-time factor running on CPU and GPU.}
    \vspace{-0.5cm}
    \label{tab:runtime}
\end{table}

\section{Conclusion}

In this work, we presented a noise reduction system the combines a signal processing spectral gating denoiser with a neural network, enabling automatic high-fidelity noise reduction for both speech and music.
We outlined the challenges in integrating this denoiser within a gradient-based machine learning paradigm and proposed a two-stage training approach that combines supervised pretraining with a gradient approximation scheme in order to facilitate efficient and stable training with the denoiser in the loop. 
We conducted both an objective evaluation along with a subjective listening test and found that our proposed approach performed on par with strong deep learning baselines as well as an industry standard noise reduction system. 
Our method achieved this while being fully automatic, interpretable, controllable, and an order of magnitude more efficient than other deep learning approaches. 
Future work could consider further improvements to the spectral gating denoiser architecture that improve perceptual performance, and low-latency operation could be achieved with the integration of time domain filterbanks.

\newpage
\section{Acknowledgment}
\label{sec:ack}
This work was supported in part by the EPSRC UKRI CDT in AI and Music (Grant no. EP/S022694/1). 

\def\bibfont{\footnotesize}

\bibliographystyle{jaes}

\bibliography{refs}

\end{document}

%% file: aestitle.tex
\savebox{\AEStop}{%
        
	\begin{minipage}{\textwidth}%
 \vspace{-1cm}
		\rule{\textwidth}{1.5pt}\\%
		\\%
		\begin{minipage}[c][\iftoggle{convention}{3.2cm}{3.7cm}][t]{0\textwidth}%
			\includegraphics[width=20mm]{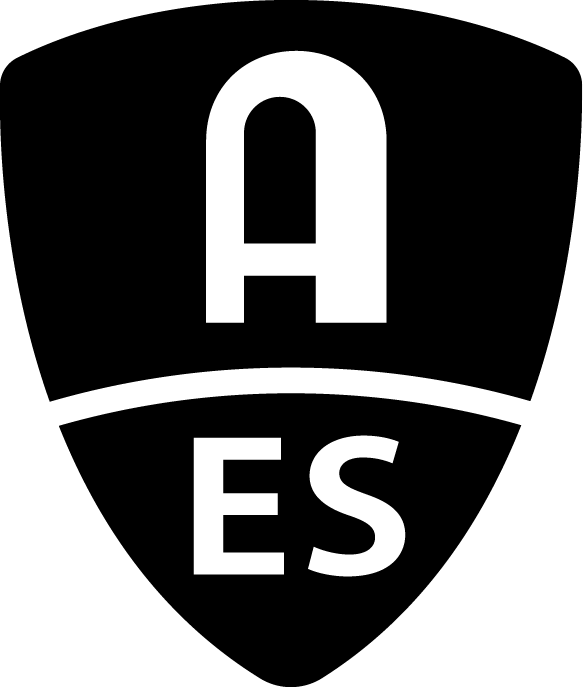}%
		\end{minipage}%
		\begin{minipage}{\textwidth}%
			\sffamily%
			\begin{center}%
				\LARGE Audio Engineering Society\\%
				\iftoggle{express_paper}{%
				\hspace{3mm}\fontsize{36}{38pt}\selectfont Convention Express\\Paper \AESExpressPaperNumber\\%
				}{%
				\iftoggle{convention}{%
				\fontsize{36}{38pt}\selectfont Convention Paper\\%
				}{%
				\fontsize{36}{38pt}\selectfont Conference Paper\\%
				}}%
				\vspace{0.2cm}%
				\large Presented at the \AESConferenceNumber \iftoggle{convention}{Convention\\}{\AESConferencePrefix Conference on\\}%
				\iftoggle{convention}{}{\AESConferenceTopic\\}%
				\AESConferenceDate\ifx\AESConferenceLocation\empty\else, \AESConferenceLocation\fi%
			\end{center}%
		\end{minipage}\\%
		\vspace{0.2cm}\\%
		\begin{minipage}{\textwidth}%
			\rmfamily\itshape\small	\AESLegalTextPrefix\ \AESLegalText%
		\end{minipage}\\%
		\\%
		\rule{\textwidth}{1.5pt}%
	\end{minipage}%
}